%% file: main.tex
\documentclass[10pt,twocolumn,letterpaper]{article}

\usepackage{multirow}
\usepackage{enumitem}
\usepackage[pagenumbers]{cvpr} 
\usepackage[pagebackref,breaklinks,colorlinks]{hyperref}
\usepackage[accsupp]{axessibility} 

\def\confName{CVPR}
\def\confYear{2024}

\title{CycleINR: Cycle Implicit Neural Representation for Arbitrary-Scale Volumetric Super-Resolution of Medical Data}

\author{Wei Fang\textsuperscript{1,2} \quad Yuxing Tang\textsuperscript{1} \quad Heng Guo\textsuperscript{1,2} \quad Mingze Yuan\textsuperscript{1,4} \quad Tony C. W. Mok\textsuperscript{1,2} \\ \quad Ke Yan\textsuperscript{1,2} \quad Jiawen Yao\textsuperscript{1,2} \quad Xin Chen\textsuperscript{3} \quad Zaiyi Liu\textsuperscript{3} \quad Le Lu\textsuperscript{1} \quad Ling Zhang\textsuperscript{1} \quad Minfeng Xu\textsuperscript{1,2}\\
\normalsize{\textsuperscript{1} DAMO Academy, Alibaba Group} \\
\normalsize{\textsuperscript{2} Hupan Lab, 310023, Hangzhou, China} \\
\normalsize{\textsuperscript{3} Guangdong Provincial People's Hospital}\\
\normalsize{\textsuperscript{4} Peking University}\\
{\tt\small lucas.fw@alibaba-inc.com}\\
}

\begin{document}
\maketitle
\input{0_abstract}
\input{1_intro}
\input{2_related_work}
\input{3_cycleINR}

\input{4_experiments}

\input{5_conclusion}

{
    \small
    \bibliographystyle{ieeenat_fullname}
    \bibliography{main}
}


\end{document}

%% file: 0_abstract.tex
\begin{abstract}
In the realm of medical 3D data, such as CT and MRI images, prevalent anisotropic resolution is characterized by high intra-slice but diminished inter-slice resolution. The lowered resolution between adjacent slices poses challenges, hindering optimal viewing experiences and impeding the development of robust downstream analysis algorithms. Various volumetric super-resolution algorithms aim to surmount these challenges, enhancing inter-slice resolution and overall 3D medical imaging quality. However, existing approaches confront inherent challenges: 1) often tailored to specific upsampling factors, lacking flexibility for diverse clinical scenarios; 2) newly generated slices frequently suffer from over-smoothing, degrading fine details, and leading to inter-slice inconsistency. In response, this study presents CycleINR, a novel enhanced Implicit Neural Representation model for 3D medical data volumetric super-resolution. Leveraging the continuity of the learned implicit function, the CycleINR model can achieve results with arbitrary up-sampling rates, eliminating the need for separate training.
Additionally, we enhance the grid sampling in CycleINR with a local attention mechanism and mitigate over-smoothing by integrating cycle-consistent loss. We introduce a new metric, Slice-wise Noise Level Inconsistency (SNLI), to quantitatively assess inter-slice noise level inconsistency. The effectiveness of our approach is demonstrated through image quality evaluations on an in-house dataset and a downstream task analysis on the Medical Segmentation Decathlon liver tumor dataset.
\end{abstract}

%% file: 1_intro.tex
\section{Introduction}
\label{sec:intro}
\begin{figure}[!ht]
    \centering 
    \includegraphics[width=0.45\textwidth]{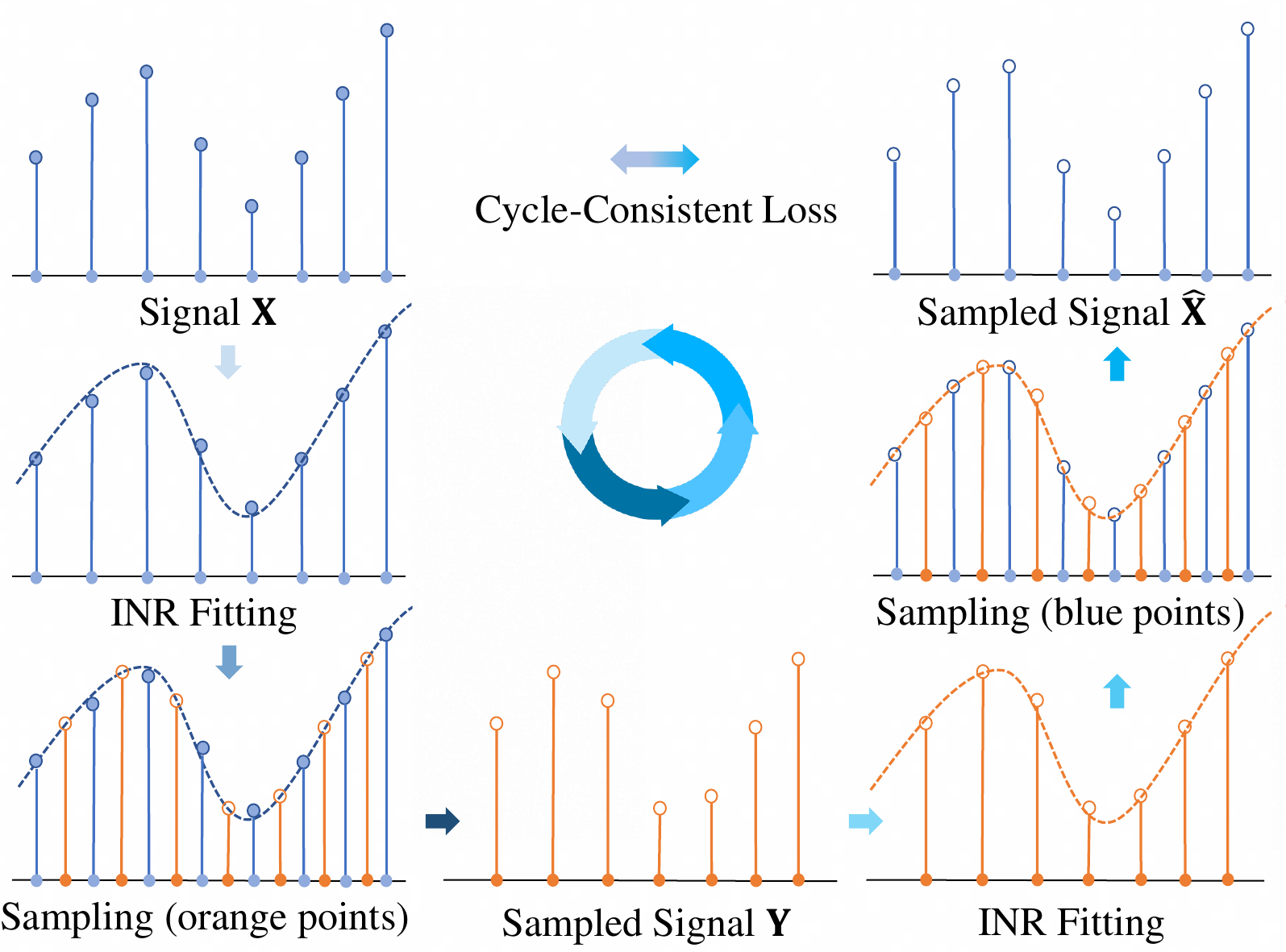} 
    \vspace{-0.2cm}
    \caption{The core concept of CycleINR involves the initial use of signal $\textbf{X}$ to fit a continuous Implicit Neural Representation (INR) function (shown in dotted blue curves). Subsequently, new points (depicted as hollow orange dots) are sampled from this function to create a new INR function (illustrated by dotted orange curves). The signal $\hat{\textbf{X}}$ is then sampled from the new function at the same positions as $\textbf{X}$. The construction of a cycle-consistent loss is achieved by assessing the similarity between $\hat{\textbf{X}}$ and $\textbf{X}$.} 
    \vspace{-0.4cm}
    \label{fig1}
\end{figure}

Volumetric medical imaging, a cornerstone of diagnostic radiology involving techniques such as computed tomography (CT) and magnetic resonance imaging (MRI), preserves the 3D three-dimensional characteristics of the body's internal structures through the compilation of multiple cross-sectional images~\cite{williams2019we}. While high-resolution volumetric medical imaging offers detailed anatomical and functional information enhancing diagnosis, its widespread clinical adoption is hindered by prolonged acquisition time and elevated storage costs. As a common alternative, anisotropic volumes, characterized by high resolution within each slice and lower resolution between slices, have become prevalent~\cite{THEVENAZ2009465}. 
However, this reduced inter-slice resolution introduces challenges: limiting detailed sagittal or coronal views, complicating lesion observation, and posing difficulties for robust 3D medical image interpretation algorithms. Addressing this, there is a critical need for an accurate 3D super-resolution method, essentially framing it as a volumetric super-resolution task~\cite{peng2021vsr, yu2022rplhr}.
It is noteworthy that video frame interpolation~\cite{reda2022film}, which involves synthesizing intermediate images between a pair of input frames, exhibits notable similarities with volumetric interpolation.

 Implementing volumetric super-resolution algorithms poses challenges on various fronts. 
 Firstly, most super-resolution methods are constrained to specific super-resolution ratios, necessitating the training of multiple models for different super-resolution levels. Secondly, the newly generated slices often exhibit over-smoothing, creating a noticeable slice-wise inconsistency issue in volumetric scenarios. This becomes evident when scrolling through the slices and observing the contrast between the original and generated slices. 
In contrast, in 2D scenarios, newly generated pixels are usually interpolated between existing pixels, making them less perceptible.

To tackle these challenges, we propose CycleINR (cycle-consistent loss enhanced implicit neural representation) as a solution for volumetric super-resolution. 
CycleINR framework employs a unified implicit neural voxel function to represent both low-resolution (LR) and high-resolution (HR) images, each at different sampling rates.
This enables treating volumetric super-resolution as an implicit neural function approximation problem. Leveraging the continuity of the implicit neural function, a single implicit voxel function achieves volumetric super-resolution with arbitrary upsampling scales.  To mitigate over-smoothing, we integrate cycle-consistent loss (CCL) into our method. This entails the utilization of pixel values at LR grid points to fit the implicit neural function, generating pixel values for initially empty points (forward path). Furthermore, we employ the newly generated HR images to reconstruct the LR images (backward path). The backward path ensures the generated image maintains a consistent noise level with the original images. These paths establish a cyclical process, ensuring both accurate and consistent HR image recovery. 
Additionally, in our sampling process from the implicit neural representation function, we design a local attention mechanism (LAM) to consider both spatial proximity and numerical similarity. This ensures accurate and reliable sampling by incorporating both local spatial relationships and pixel value similarities.
Finally, we introduce a novel metric, slice-wise noise level inconsistency (SNLI), to quantitatively measure inter-slice inconsistency, tackling limitations in existing metrics such as 
PSNR (peak signal-to-noise ratio), SSIM (structural similarity index), and LPIPS (learned perceptual image patch similarity)~\cite{zhang2018unreasonable}. These metrics are more favorable towards assessing the smoothness or single-image similarity, making SNLI crucial for evaluating inter-slice variations in volumetric data and showcasing the effectiveness of CCL.

Our contributions are outlined as follows:
\begin{enumerate}[leftmargin=*, noitemsep]
\item We introduce CycleINR, a novel framework for volumetric super-resolution in 3D medical imaging. Leveraging its coordinate-based nature, our method offers flexibility for arbitrary upsampling scales and the ability to generate multiple super-resolution results without the need for additional training.
\item We propose a cycle-consistent loss within the implicit neural representation framework to address the slice-wise inconsistency problem, ensuring consistency between newly generated and existing images. Additionally, a local attention mechanism captures both spatial proximity and numerical similarity, enhancing the depiction of relationships between pixels.
\item We propose a new metric SNLI (slice-wise noise level inconsistency) to quantitatively measure inter-slice noise level inconsistency in 3D data.
\item We conduct extensive experiments, evaluating both image quality and downstream tasks, demonstrating the effectiveness of our proposed method.
\end{enumerate}

%% file: 2_related_work.tex
\section{Related Work }
\label{sec:related_work}

\begin{figure*}[!t]
    \centering 
    \includegraphics[width=1.0\textwidth]{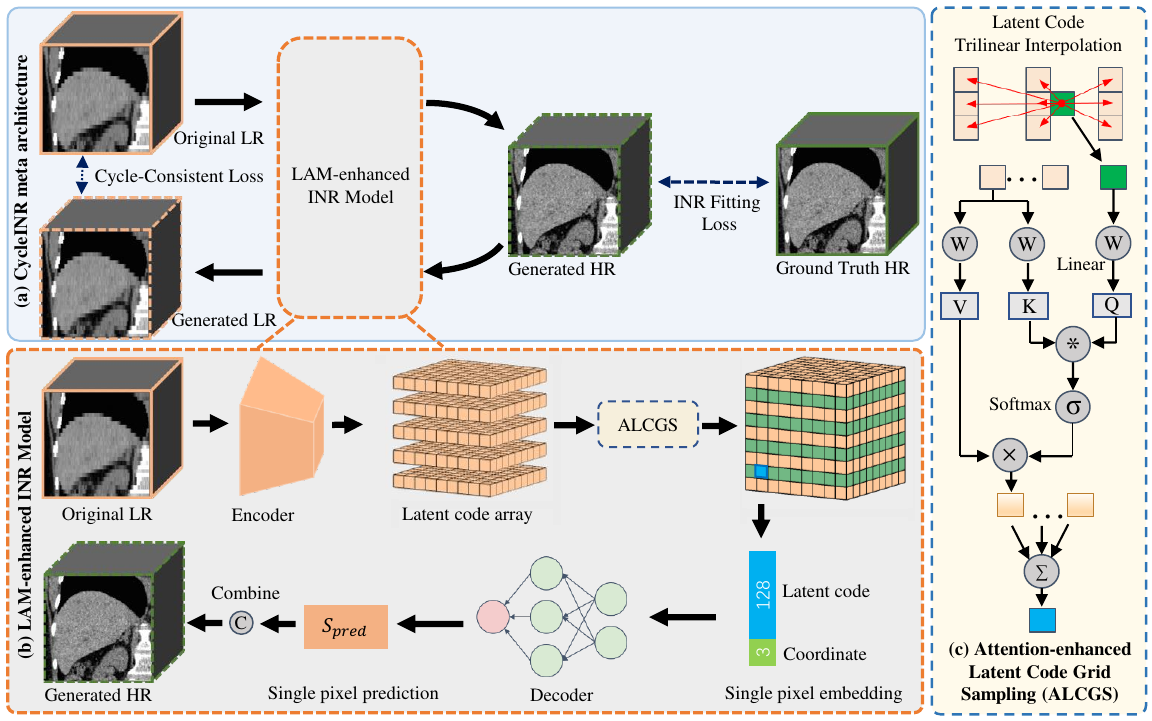} 
    \vspace{-0.4cm}
    \caption{Schematic plot of our CycleINR framework for volumetric super-resolution. (a) Meta architecture of the proposed CycleINR. (b) Local Attention Mechanism (LAM) enhanced INR model. (c) Attention-enhanced Latent Code Grid Sampling (ALCGS) process.}
    \vspace{-0.3cm}
    \label{fig2}
\end{figure*}


\quad
\textbf{Single Image Super-Resolution.} Single image super-resolution (SISR) has garnered significant attention in computer vision, evolving over decades~\cite{yang2019deep}. 
Common interpolation-based methods like cubic and linear interpolation are widely employed for their simplicity~\cite{blu2004linear}. Image prior like total variation can also enhance image super-resolution by formulating the task as a convex optimization problem~\cite{shi2015lrtv}. In the era of deep learning, Dong et al.~\cite{dong2015image} introduced super-resolution convolutional neural networks (CNN) to learn the mapping from LR images to HR images. Subsequent SISR research has delved into refining deep learning models, exploring strategies like deeper network architectures~\cite{kim2016accurate}~\cite{fang2022cross}, recursive-supervision~\cite{kim2016deeply}, and deep learning as priors~\cite{zhang2017learning}. Advanced upsampling methods in deep learning include deconvolution for direct HR image generation~\cite{dong2016accelerating}, sub-pixel convolution, and pixel shuffling for efficient upsampling at the output stage~\cite{shi2016real}. Residual learning~\cite{lim2017enhanced} and the integration of generative adversarial networks~\cite{ledig2017photo, wang2018esrgan, wang2021real} have further enriched the SISR landscape. 

\textbf{Implicit Neural Representation}. Digitalized visual signals transition from their natural continuous form to discrete representations, such as 2D images or volumetric medical data. Implicit neural representation (INR) emerges as a coordinate-based strategy to address this transformation, employing a trainable continuous function often implemented with a straightforward fully connected network~\cite{sitzmann2020implicit}.  
INR was first applied in 3D shape modeling~\cite{chen2019learning, park2019deepsdf, mescheder2019occupancy}. Subsequently, it has been extensively employed across various 3D tasks, including object shape modeling~\cite{genova2020local, atzmon2020sal}, scene reconstruction~\cite{sitzmann2019scene, jiang2020local}, and structure rendering\cite{niemeyer2020differentiable, liu2020neural}. 
Notably, INR has made significant contributions to view synthesis, exemplified by the neural radiance fields (NeRF)~\cite{mildenhall2021nerf}. INR has also been applied to 2D image super-resolution~\cite{xu2021ultrasr, liu2021enhancing, yang2021implicit, lee2022local, chen2023cascaded}. Chen et al.~\cite{chen2021learning} used INR to learn a continuous function for 2D images. The function, taking an image coordinate and surrounding 2D deep features as inputs, predicts the RGB value at the given coordinate, enabling image generation at arbitrary resolutions. Sitzmann et al.~\cite{sitzmann2020implicit} enhanced image representation by replacing the rectified linear units (ReLU) with periodic activations in multi-layer perceptron, achieving higher image quality.

\textbf{Volumetric Super-Resolution.} Volumetric super-resolution in 3D volumetric medical images, which aims to enhance spatial resolution between slices, has witnessed significant progress. DA-VSR~\cite{peng2021vsr} addressed domain adaptation challenges, SAINT~\cite{peng2020saint} employed a two-stage framework with separate 2D CNNs for sagittal and coronal upscaling, ArSSR~\cite{wu2022arbitrary} offered an INR-based arbitrary scale SISR model for 3D MR images, and TVSRN~\cite{yu2022rplhr} introduced a transformer-based approach for CT scans. Additionally, Fang et al.~\cite{fang2022incremental} proposed a medical slice synthesis method using mutual distillation.

%% file: 3_cycleINR.tex
\section{CycleINR}
\label{cycleINR}


Volumetric super-resolution can be viewed as a z-axis upsampling process applied to 3D volume data. Let $\mathbf{I^{lr}}\left(x, y, z\right) \in \mathbb{R}^{X\times Y \times Z}$ denote the original low-resolution (LR) CT scan, where $X$, $Y$, and $Z$ represent the number of slices along the x, y, and z axes, respectively. The desired output of volumetric super-resolution is denoted as $\mathbf{I^{\text{vsr}}}\left(x, y, z\right)\in \mathbb{R}^{X \times Y \times[Z\cdot r]}$, where $r>1$ represents the upsampling factor along the z-axis. This implies that super-resolution is applied solely in the z-direction, maintaining the original resolution in the x and y directions, allowing more detailed information between slices in the volumetric data. For instance, setting the upsampling factor $r$ to 5 facilitates the conversion from 5mm slice-thickness data to 1mm, or 4 for the conversion from 5mm to 1.25mm slice-thickness. In the context of super-resolution methods employing the coordinate-based INR (implicit neural representation) model, $r$ can also be a decimal, given the availability of coordinates. The volumetric super-resolution transform can be represented as:
\begin{equation}
\label{eq1}
\mathcal{F}: \mathbf{I^{lr}}\in\mathbb{R}^{X\times Y \times Z}\to \mathbf{I^{\text{vsr}}}\in\mathbb{R}^{X\times Y \times [Z\cdot r ]}.
\end{equation}

\subsection{Local Attention Mechanism-Enhanced INR}
In the original INR model, 3D volumetric medical data can be represented using a coordinate-based implicit voxel function $I=f_\theta(\mathbf{x})$,
where $\theta$ represents the trainable parameters of the INR model, $\mathbf{x} = (x, y, z)$ denotes any 3D voxel spatial coordinate, and $I$ denotes the voxel intensity at the coordinate $\mathbf{x}$ in the image. The spatial coordinates are normalized to the range [-1, 1] along the x, y, and z dimensions, differing from physical coordinates. 
Both HR and LR images can be regarded as the explicit representation of the implicit function $f_\theta(x)$ but at different sampling rates. Once the voxel function $f_\theta(x)$ is well approximated, the desired super-resolution results at an arbitrary scale can be achieved by simply adjusting the sampling rate (Figure~\ref{fig1}).

However, in $I = f_\theta(\mathbf{x})$, the image information is decoded into the parameters of the function, restricting the function's specificity to the represented image. As training a unique model for each image is impractical, a more flexible function that can represent multiple images is essential. Inspired by~\cite{chen2021learning}, we reformulate the voxel function by incorporating a latent code using $I=f_\theta(\mathbf{x},\mathbf{z})$,
where $\mathbf{z}$ is a vector that represents the latent code capturing pixel-specific spatial features, which is generated for each pixel through a convolutional encoder, capturing local semantic information around the pixels. Hence, it can be considered as a latent representation for the pixel in the latent space. The architecture of the INR model with local attention mechanism enhanced latent code grid sampling process, is illustrated in Figure \ref{fig2} (b). The following paragraphs will delve into a detailed explanation of these designs.

\textbf{Encoder Network.}
We first employ an encoder $\mathcal{E_{\theta}}$ to extract semantic features from the LR volume. The encoder takes the LR image volume $\mathbf{I}^{lr} \in \mathbb{R}^{h\times w \times d}$ as input and produces a feature map $\mathbf{Z}^{lr} \in \mathbb{R}^{h\times w \times d \times |\mathbf{z}|}$, referred to as the latent code. This implies that each voxel in the LR volume is transformed into a latent code with a length of $|\mathbf{z}|$. The encoding process can be represented as:
\begin{equation}
\label{eq4}
\mathbf{Z}^{lr} = \mathcal{E_{\theta}}(\mathbf{I}^{lr}),
\end{equation}
where $\theta$ is the trainable parameters of the encoder. We utilize a residual convolutional neural network encoder as described in \cite{du2020super}.
Our strategy to extract voxel-wise latent space is crucial for facilitating the decoder network in seamlessly incorporating local image intensity information. This integration proves instrumental in effectively recovering fine details within the high-resolution image, especially when dealing with large upsampling scales.

\textbf{Attention-enhanced Latent Code Grid Sampling.}
Following the encoding process, the obtained latent code corresponds to the voxels in the LR volume.
To acquire the latent code for the voxels at coordinate $\mathbf{x}$ in the HR volume $\mathbf{Z}^{hr}$, an additional operation known as grid sampling is required. Prior works~\cite{chen2021learning, wu2022arbitrary} performed grid sampling using techniques such as trilinear interpolation or cell decoding, which can be expressed as:
\begin{equation}
\label{eq5}
\mathbf{\hat{Z}}_{\mathbf{x}}^{hr} = \sum_{k=1}^{L}(\frac{S_k}{S}\cdot \mathbf{Z}_k^{lr}{}),
\end{equation}
where $\mathbf{Z}_k^{lr} (k=1,...,L)$ denotes the latent code of the $L=8$ surrounding coordinates in the feature map $\mathbf{Z}^{lr}$. $S_k$ represents the volume of the cuboid formed using the coordinate $\mathbf{x}$ and the coordinate diagonal to the $k_{th}$ coordinate, and $S$ is the sum of all $S_k$.

To better depict the relationships between voxels, we propose an attention-enhanced latent code grid sampling (ALCGS) mechanism
to enhance the generated voxel latent code at coordinate $\mathbf{x}$, which is represented as follows:
\begin{equation}
\label{eq6}
\mathbf{\Tilde{Z}}_{\mathbf{x}}^{hr} = \sum_{i=1}^{N} \sigma (\mathcal{W}_{Q} (\mathbf{\hat{Z}}_{\mathbf{x}}^{hr}) \cdot \mathcal{W}_{K}(\mathbf{Z}_i^{lr}))\mathcal{W}_{V}(\mathbf{Z}_i^{lr}),
\end{equation}
where $N$ is the number of neighboring latent codes used for ensembling, and $\sigma$ denotes the Softmax function. Specifically, the initial latent code $\mathbf{\hat{Z}}_{\mathbf{x}}^{hr}$ of the voxel at coordinate $\mathbf{x}$ is firstly computed by using trilinear interpolation. Subsequently, $N$ nearest neighboring latent codes $\mathbf{Z}_i^{lr}$ are selected around $\mathbf{\hat{Z}}_{\mathbf{x}}^{hr}$ for aggregation. The initial latent code $\mathbf{\hat{Z}}_{\mathbf{x}}^{hr}$ is then mapped to the query space, while the $N$ nearby latent codes $\mathbf{Z}_i^{lr}$ are mapped to the key and value space, respectively, through linear layers $\mathcal{W}_Q$, $\mathcal{W}_K$, and $\mathcal{W}_V$. Following this, the similarities between the query vector and each key vector are computed, which are then converted to weight coefficients through the Softmax function. Finally, the improved latent code $\mathbf{\Tilde{Z}}_{\mathbf{x}}^{hr}$ is obtained by weighted summation.

The local attention mechanism establishes pixel relationships not solely dependent on physical distance but also on visual information and latent code similarity. This aggregation strategy enhances trilinear interpolation and enables the generated latent codes to preserve richer details, similar to the functionality of bilateral filtering. 

\textbf{Decoder Network.}
The decoder network $ \mathcal{D}_\theta$ takes the query voxel coordinate $\mathbf{x}$ and its latent code  $\mathbf{\Tilde{Z}}_{\mathbf{x}}^{hr}$ as input and outputs the estimated voxel value $I_{hr}(\mathbf{x})$, expressed as:
\begin{equation}
\label{eq7}
I^{hr}(\mathbf{x}) = \mathcal{D}_\theta(\mathbf{x},\mathbf{\Tilde{Z}}_{\mathbf{x}}^{hr}).
\end{equation}
Following the design of \cite{wu2022arbitrary}, the decoder network comprises eight standard fully connected (FC) layers, with a ReLU activation following each FC layer. A residual connection is employed between the input of the decoder network and the intermediate fourth FC layer.

\subsection{Cycle-Consistent Loss for INR Model}
Inspired by CycleGAN~\cite{zhu2017unpaired}, we integrate a cycle-consistent loss into the INR model. This loss acts as a constraint to ensure consistency between the generated image and the original image in terms of image features, such as noise level. Despite both being generative models, generative adversarial networks (GANs) and INR diverge in their objectives. GANs transform one probability distribution function into another, whereas the INR model learns a continuous function from discrete data. 
Arbitrary scaled super-resolution images can be sampled from this continuous function. The newly sampled image can also be used to fit a new continuous function.
We assert that the newly generated function should remain consistent with the original discrete data, leading to the incorporation of cycle-consistent loss in the INR model. 

The synergy between cycle-consistent loss and the INR model is highly effective, highlighting the INR model's superior suitability for incorporating cycle-consistent loss compared to GANs. GANs are designed to transform between distinct input and output distributions, while INR models aim to model the same function for both the input LR image and the output HR image. It is expected that the newly generated continuous function should produce values matching the original discrete data when sampled. 

Let's represent the local attention mechanism (LAM)-enhanced INR model described in Figure \ref{fig2} (b) as $\mathcal{G_{\theta}}$. In this context, the cycle-consistent loss is articulated as:
\begin{gather}
\label{eq8}
\mathbf{\hat{I}}_{j}^{hr}=\mathcal{G_{\theta}}(\mathbf{I}_{j}^{lr},\mathbf{C}_{j}^{hr}), \\ 
\label{eq9}
\mathcal{L}_{Cycle} = \sum_{j}^{M}|| \mathcal{G_{\theta}}(\mathbf{\hat{I}}_{j}^{hr}, \mathbf{C}_{j}^{lr})-\mathbf{I}_{j}^{lr}||_{1},
\end{gather}
where $j$ and $M$ are the index and the number of training samples, respectively. The INR model $\mathcal{G_{\theta}}$ takes the LR image $\mathbf{I}_{j}^{lr}$ and the coordinates of the voxels to be generated $\mathbf{C}_{j}^{hr}$ as input, producing the sampled HR image $\mathbf{\hat{I}}_{j}^{hr}$. In Equation \ref{eq9}, the newly generated image $\mathbf{\hat{I}}_{j}^{hr}$, along with the coordinates of the original LR image $\mathbf{C}_{j}^{lr}$, is once again fed as input into the INR model to generate an estimation for the LR image. The estimated LR image is subsequently compared for similarities with the original LR image $\mathbf{I}_{j}^{lr}$ using the L1 loss, constituting the cycle-consistent loss. 

It is worth noting that HR pixel coordinates may not cover all LR pixel coordinates, given the random sampling of scaling factors, which may involve decimals during training. Even with integers, especially when a scaling factor is an even number and the interpolation mode align-corner is set to False, there may be no overlap between the pixel positions in the upscaled image and those of the original LR image. This complexity makes the downsampling process in Equation \ref{eq9} non-trivial, and obtaining it can not be accomplished by simply selecting pixels at regular intervals.

The forward INR fitting loss can be represented as:
\begin{equation}
\label{eq10}
\mathcal{L}_{INR} = ||\mathbf{\hat{I}}_{j}^{hr}-\mathbf{I}_{j}^{hr}||_{1}.
\end{equation}

The overall loss of training is represented as follows:
\begin{equation}
\label{eq11}
\mathcal{L}_{CycleINR} = \mathcal{L}_{INR} + \lambda \cdot \mathcal{L}_{Cycle},
\end{equation}
where $\lambda$ serves as a regularization factor that helps to balance the INR fitting loss and cycle-consistent loss.

\begin{figure*}[!t]
    \includegraphics[width=1.0\textwidth]{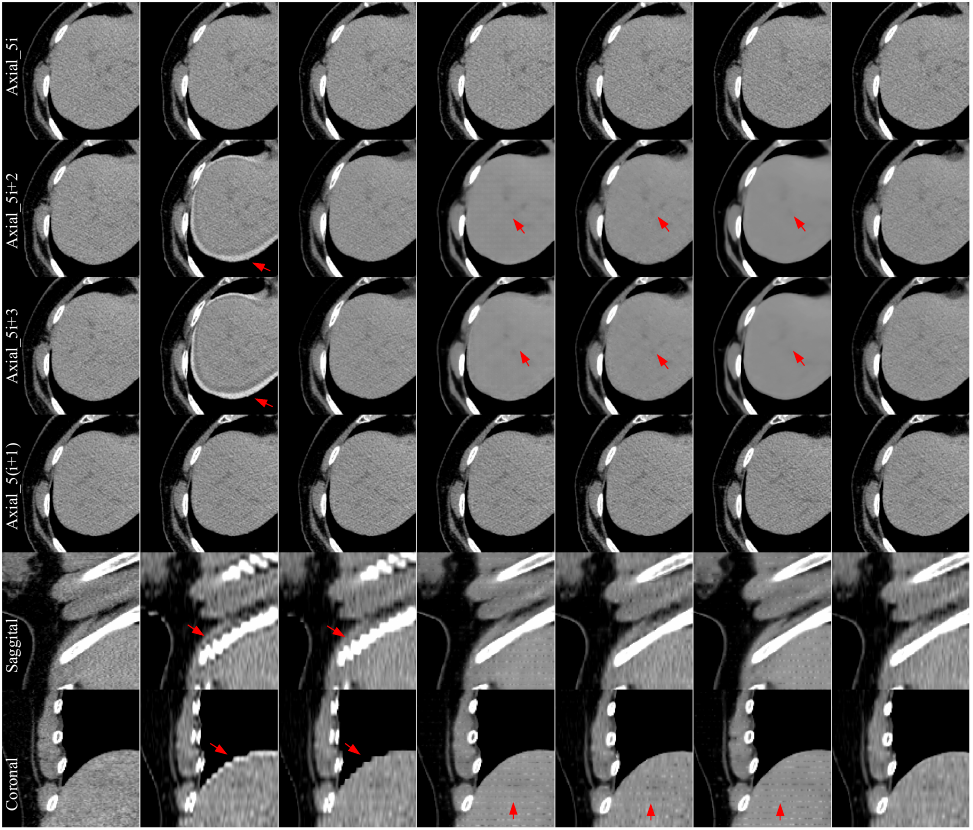} 
    \begin{minipage}{\textwidth}
        \begin{tabular}{p{2.1cm}p{2.1cm}p{2.1cm}p{2.1cm}p{2.1cm}p{2.1cm}p{2.1cm}}
         \centering GT  & \centering Cubic & \centering Trilinear & \centering TVSRN\cite{yu2022rplhr} & \centering ArSSR\cite{wu2022arbitrary} & \centering SAINT\cite{peng2020saint} & \centering CycleINR \\
        \end{tabular}
    \end{minipage}
    \setlength{\abovecaptionskip}{0pt}
    \vspace{-0.3cm}
    \caption{Visualization results of different super-resolution (x5) methods from axial, sagittal, and coronal views. The second and third rows display newly generated images at new positions, while the first and fourth rows show generated images at their original positions. In the bottom row, red arrows in TVSRN, ArSSR, and SAINT highlight horizontal lines in the coronal view, indicating slice-wise inconsistency.}
    \vspace{-0.3cm}
    \label{fig3}
\end{figure*}

\subsection{SNLI: Measure Slice-wise Noise Inconsistency}
To quantitatively measure the inter-slice inconsistency caused by the over-smoothing in the newly generated slices, we propose a novel metric named slice-wise noise level inconsistency (SNLI). This inconsistency often manifests as a perceptible `slice discontinuity' phenomenon during axial slice scrolling. Notably, there is currently no existing metric that adequately captures this phenomenon. The proposed SNLI metric is defined as follows:
\begin{equation}
\label{eq12}
\text{SNLI} = \Psi({\zeta(\textbf{I}_{t})}), t=0,1,...,T,
\end{equation}
where $\textbf{I}_{t} \in \mathbb{R}^{X\times Y}$ is the $t_{th}$ axial slice of the image volume. $\zeta$ is a robust wavelet-based estimator of the Gaussian noise standard deviation, described in \cite{donoho1994ideal}. $\Psi$ represents the process of calculating the standard deviation. Thus, the SNLI metric helps measure the slice-wise noise level inconsistency between axial slices, which was ignored by many deep learning-based volumetric super-resolution methods.

%% file: 4_experiments.tex
\section{Experiments}
\label{Experiments}

\begin{table*}[!ht]
\centering
\begin{tabular}{ccp{2cm}p{1.4cm}cccc}
\toprule
Scale   & Method & PSNR($\uparrow$) & SSIM($\uparrow$) & LPIPS\_alex($\downarrow$)   & LPIPS\_vgg($\downarrow$) & LPIPS\_squeeze($\downarrow$) & SNLI($\downarrow$) \\
\midrule
\multirow{6}{*}{x2} 
& Cubic          & 39.5039 & 0.9705 & \textbf{0.0148} & 0.0639 & \textbf{0.0161} & \underline{0.5318} \\
& Trilinear      & 40.6750 & 0.9757 & \underline{0.0201} & \textbf{0.0618} & \underline{0.0189} & 0.8450 \\
& TVSRN \cite{yu2022rplhr}         & \underline{43.6167} & \underline{0.9820} & 0.0311 & 0.0824 & 0.0310 & 1.4307 \\
& ArSSR \cite{wu2022arbitrary}         & 42.6713 & 0.9799 & 0.0370 & 0.0913 & 0.0357 & 1.5738 \\
& SAINT \cite{peng2020saint}          & \textbf{44.3977} & \textbf{0.9833} & 0.0361 & 0.0867 & 0.0354 & 1.6724 \\
& CycleINR (Ours) & 43.0137 & 0.9805 & \underline{0.0201} & \underline{0.0625} & 0.0206 & \textbf{0.3527} \\
\midrule
\multirow{6}{*}{x3} 
& Cubic          & 35.0140 & 0.9393 & \underline{0.0306} & 0.1084 & \underline{0.0292} & \textbf{0.5322} \\
& Trilinear      & 36.5867 & 0.9515 & 0.0328 & \underline{0.0973} & 0.0293 & 0.7422 \\
& TVSRN \cite{yu2022rplhr}         & \underline{40.4857} & \underline{0.9696} & 0.0607 & 0.1419 & 0.0583 & 1.4389 \\
& ArSSR \cite{wu2022arbitrary}         & 39.3398 & 0.9659 & 0.0443 & 0.1135 & 0.0413 & 1.3026 \\
& SAINT \cite{peng2020saint}          & \textbf{40.8705} & \textbf{0.9711} & 0.0625 & 0.1404 & 0.0599 & 1.7535 \\
& CycleINR (Ours) & 39.2748 & 0.9644 & \textbf{0.0293} & \textbf{0.0902} & \textbf{0.0280} & \underline{0.6674} \\
\midrule
\multirow{6}{*}{x5} 
& Cubic          & 31.0470 & 0.8896 & 0.0562 & 0.1627 & 0.0480 & \textbf{0.5336} \\
& Trilinear      & 32.6606 & 0.9106 & \underline{0.0525} & \underline{0.1413} & \underline{0.0433} & 0.6682 \\
& TVSRN \cite{yu2022rplhr}         & \underline{36.8459} & \underline{0.9503} & 0.0927 & 0.1989 & 0.0862 & 1.3593 \\
& ArSSR \cite{wu2022arbitrary}          & 35.1960 & 0.9394 & 0.0611 & 0.1485 & 0.0528 & 1.1714 \\
& SAINT \cite{peng2020saint}         & \textbf{36.9940} & \textbf{0.9519} & 0.0996 & 0.2044 & 0.0951 & 1.6996 \\
& CycleINR (Ours) & 35.0022 & 0.9354 & \textbf{0.0464} & \textbf{0.1289} & \textbf{0.0399} & \underline{0.6050} \\
\bottomrule
\end{tabular}

\caption{Image quality evaluation of different methods. \textbf{bald} signifies the best, and \underline{underline} represents the suboptimal. Note that PSNR and SSIM are not the sole determinants, and SNLI proves particularly valuable for comparing deep learning-based methods.}
\label{tab1}
\end{table*}

\subsection{Experimental Setup}
\quad
\textbf{Data.} We employed two datasets in our experiment. The first, a privately collected chest CT data (referred to as the chest dataset hereafter) consists of 204 CT volumes with 1mm slice thickness. We randomly split it into 124 for training, 40 for validation, and 40 for testing. 
Following established practices in volumetric super-resolution~\cite{wu2022arbitrary, peng2020saint}, we derived the LR volumes by down-sampling from the corresponding HR volume, creating LR-HR training pairs. The CycleINR model and other comparing learning-based methods were trained on these pairs. Testing involved downsampling to 2mm, 3mm, and 5mm thickness using interval sampling, aiming to recover 1mm data using super-resolution approaches. Image quality was evaluated by comparing results with ground truth 1mm data. The second dataset is from the public Medical Segmentation Decathlon \cite{antonelli2022medical} Task03\_Liver training set (referred to as the MSD liver dataset). It serves to assess whether the HR images generated by the CycleINR model achieve performance comparable to the original HR images in downstream segmentation tasks. The MSD liver dataset consists of 131 volumes with liver and tumor masks. For evaluation, we selected 33 volumes with 1mm slice thickness as the test set and utilized the remaining 98 image-mask pairs to train a segmentation model using nnUNet~\cite{isensee2021nnu}. The test data was downsampled and then restored to 1mm as on the chest CT dataset. We compared the segmentation results of different methods, setting the segmentation from the original 1mm data as references. 

\textbf{Implementation Details.} The encoder consists of three blocks, where each block includes six 3D convolutional layers followed by ReLU. The latent code vector length $|\mathbf{z}|$ is set to 128. The feature channels of the intermediate layers are set to 64. The number of neighboring latent codes $N$ used for ensembling is set to 27 to include all the $3 \times 3 \times 3$ local regions around a specific voxel. The decoder is configured with 8 fully connected layers, where the intermediate layer has a dimension of 256. During CycleINR training, LR-HR pairs are randomly sampled on the fly from the original entire volume with an upsample scale ranging from 2.0 to 6.0, with one decimal point precision. The balancing factor $\lambda$ for the INR fitting loss and cycle-consistent loss is set to 1. The Adam optimizer \cite{kingma2014adam} minimizes the total loss in Equation \ref{eq11}, with an initial learning rate of $10^{-4}$, and decays as half every 200 epochs. The training runs for a total of 3,000 epochs. During testing, Gaussian window-weighted sliding window inference is employed to mitigate stitching artifacts. Note that these hyper-parameters are set following empirical observations or prior works.

\textbf{Compared Methods.} 
For comparison, we include cubic interpolation, trilinear interpolation, and three additional deep learning-based volumetric super-resolution methods: TVSRN~\cite{yu2022rplhr}, ArSSR~\cite{wu2022arbitrary}, and SAINT~\cite{peng2020saint}. These methods, featuring distinct architectural designs, demonstrate the capability of achieving arbitrary upscaling ratios, except for TVSN, which requires training separate models for super-resolution factors of x2, x3, and x5, respectively. 

\begin{table*}[t]
\centering
\begin{tabular}{ccccccccc}
\toprule
 INR &LAM &CCL & PSNR($\uparrow$) & SSIM($\uparrow$) & LPIPS\_alex($\downarrow$)   & LPIPS\_vgg($\downarrow$) & LPIPS\_squeeze($\downarrow$) & SNLI($\downarrow$) \\
\midrule
 
\checkmark & &          & \textbf{35.1960} & \textbf{0.9394} & 0.0611 & 0.1485 & 0.0528 & 1.1714 \\
\checkmark & \checkmark &        & \underline{35.1395} & \underline{0.9391} & 0.0603 & 0.1478 & 0.0522 & 1.1395 \\
\checkmark & & \checkmark     & 34.9610 & 0.9350 & \underline{0.0470} & \underline{0.1298} & \underline{0.0405} & \underline{0.6134} \\
\checkmark & \checkmark & \checkmark   & 35.0022 & 0.9354 & \textbf{0.0464} & \textbf{0.1289} & \textbf{0.0400} & \textbf{0.6050} \\
\bottomrule
\end{tabular}
\vspace{-0.2cm}
\caption{Ablation study on the Local Attention Mechanism (LAM) and Cycle-Consistent Loss (CCL).}
\vspace{-0.21cm}
\label{tab3}
\end{table*}

\textbf{Evaluation Metrics.} For the image quality evaluation, we first use peak signal-to-noise ratio (PSNR) and structural similarity (SSIM)~\cite{wang2004image} as fundamental metrics for different super-resolution methods. However, acknowledging the bias of PSNR and SSIM towards favoring blurry images, where an over-smoothed super-resolution result could yield high scores under these metrics despite significant loss of image details, we also include learned perceptual image patch similarity (LPIPS)~\cite{zhang2018unreasonable} in the evaluation. LPIPS scores are calculated using three backbones (AlexNet~\cite{krizhevsky2012imagenet}, VGGNet~\cite{simonyan2014very}, and SqueezeNet~\cite{iandola2016squeezenet}). In addition, we calculate the SNLI metric to measure the inconsistency between slices, often ignored by most volumetric super-resolution methods. For downstream segmentation task evaluation, we use the Dice similarity coefficient (DSC)~\cite{dice1945measures} and normalized surface Dice (NSD)~\cite{nikolov2018deep} with a tolerance of 1mm.
\begin{table}[t]
\centering
\begin{tabular}{p{0.1cm}cp{0.9cm}p{0.9cm}p{0.9cm}p{0.95cm}}
\toprule
\small{Scale}   & \small{Method} & \small{DSC\_L$\uparrow$} & \small{NSD\_L$\uparrow$} & \small{DSC\_T$\uparrow$}   & \small{NSD\_T$\uparrow$}  \\
\midrule
\multirow{6}{*}{x2} 
& Cubic              & 0.9868 & 0.9568 & \textbf{0.9069} & \textbf{0.8632}   \\
& Trilinear          & \underline{0.9892} & \underline{0.9622} & \textbf{0.9069} & \underline{0.8575}   \\
& TVSRN \cite{yu2022rplhr}             & 0.9886 & 0.9514 & 0.8552 & 0.8065  \\
& ArSSR \cite{wu2022arbitrary}             & 0.9856 & 0.9459 & 0.8516 & 0.7952  \\
& SAINT \cite{peng2020saint}             & 0.9884 & 0.9540 & 0.8476 & 0.7904  \\
& CycleINR           & \textbf{0.9911} & \textbf{0.9648} & 0.8730 & 0.8226  \\

\midrule
\multirow{6}{*}{x3} 
& Cubic              & 0.9625 & 0.7862 & 0.7621 & 0.6891   \\
& Trilinear          & 0.9783 & 0.9037 & 0.7858 & 0.7083   \\
& TVSRN \cite{yu2022rplhr}            & 0.9783 & 0.9105 & 0.7672 & 0.6892  \\
& ArSSR \cite{wu2022arbitrary}             & 0.9777 & 0.9056 & 0.7811 & 0.6988  \\
& SAINT \cite{peng2020saint}             & \underline{0.9820} & \underline{0.9250} & \underline{0.8042} & \underline{0.7338}  \\
& CycleINR           & \textbf{0.9832} & \textbf{0.9294} & \textbf{0.8087} & \textbf{0.7381}  \\

\midrule
\multirow{6}{*}{x5} 
& Cubic              & 0.8678 & 0.4475 & 0.5644 & 0.4386   \\
& Trilinear          & 0.9487 & 0.6873 & \underline{0.6755} & 0.5385   \\
& TVSRN \cite{yu2022rplhr}             & 0.9481 & 0.8228 & 0.6440 & \underline{0.5431}   \\
& ArSSR \cite{wu2022arbitrary}             & 0.9589 & 0.7907 & 0.6582 & 0.5297   \\
& SAINT \cite{peng2020saint}             & \underline{0.9619} & \underline{0.8561} & 0.6006 & 0.5050   \\
& CycleINR           & \textbf{0.9620} & \textbf{0.8566} & \textbf{0.6881} & \textbf{0.5640}   \\
\bottomrule
\end{tabular}
\vspace{-0.2cm}
\caption{Downstream task evaluation on the MSD liver tumor dataset. Comparison of segmentation across different super-resolution methods with regard to the segmentation on the original 1mm data. '\_L' and '\_T' represent the liver and tumor, respectively. Results are compared to segmentation on the original 1mm image, not the ground truth mask, to evaluate the fidelity of the upscaled image to the original 1mm characteristics.}

\label{tab2}
\end{table}
\subsection{Super-resolution Results Evaluation}
\quad
\textbf{Image Quality Evaluation.} Figure \ref{fig3} visualizes the super-resolution results produced by different methods, and Table \ref{tab1} presents the corresponding image quality metrics. In summary, traditional cubic and trilinear interpolation methods exhibit limitations in recovering HR images, particularly at large upscaling scales such as x5, where bone structures are notably challenging to recover, resulting in low PSNR and SSIM scores. While the three deep learning-based methods demonstrate relatively favorable results, a shared challenge is their oversight of over-smoothness in the generated slices, as evident in both the visualization results and the SNLI metric. In Figure \ref{fig3}, the newly generated axial slices are over-smoothed, leading to the loss of fine vessel details in the liver. Moreover, in the sagittal and coronal views, horizontal lines emerge due to noise level inconsistencies between slices. The SNLI metric for these three deep learning-based methods is markedly elevated compared to other methods. However, with the enhancement of CCL, CycleINR demonstrates significant improvements, effectively recovering fine details without over-smoothing in the newly generated slices, and revealing the absence of visible horizontal lines in sagittal and coronal views. Despite a slight decrease in PSNR and SSIM scores, it is important to note, as mentioned in \cite{zhang2018unreasonable}, that this decline does not necessarily signify a negative outcome. Furthermore, the SNLI metric demonstrates the superiority of the proposed CycleINR in preserving consistency between axial slices compared to other deep learning-based methods. This emphasizes the substantial effectiveness of employing CCL in mitigating over-smoothing issues in newly generated slices.

\textbf{Downstream Task Evaluation.} To make a comprehensive comparison of the super-resolution results generated by various deep learning models, we adopt a segmentation evaluation strategy, aligning with approaches from prior works~\cite{peng2023deep, wang2020enhanced}. This strategy allows for a thorough assessment of the SR results, moving beyond visual appearance to provide a more clinically relevant measure of performance. As depicted in Table \ref{tab2}, the segmentation results of CycleINR align better with the reference compared to other methods, particularly superior at higher upsampling scales. 
While traditional interpolation methods might retain advantages in specific scenarios with smaller upscaling ratios (e.g., x2), it is crucial to highlight that our proposed CycleINR consistently exhibits superior performance compared to other deep learning methods across various upscaling ratios. This underscores the clinical relevance of our method and establishes its efficacy in achieving better results for downstream tasks.

\subsection{Ablation Studies}
We perform ablation studies to assess the impact of two key components in our CycleINR framework: the local attention mechanism (LAM) during latent code grid sampling, and the cycle-consistent loss (CCL) aimed at minimizing inter-slice consistency. The contributions of these components to the overall performance are shown in Table \ref{tab3}.

\textbf{Local Attention Mechanism (LAM).} 
The local attention mechanism in our process effectively combines physical distance and latent code similarities to aggregate weights for latent code representation. Drawing inspiration from bilateral filtering, our LAM employs attention mechanisms for improved expressive power. As shown in Table \ref{tab3}, integrating LAM into the latent code grid sampling process enhances the LPIPS metric~\cite{zhang2018unreasonable} by preserving edges and mitigating excessive smoothing, similar to the principles of traditional bilateral filtering \cite{tomasi1998bilateral}.

\textbf{Cycle-Consistent Loss (CCL).} The purpose of CCL is to impose a constraint ensuring consistency between the generated and original images in terms of image characteristics, such as noise level. As shown in Table \ref{tab3}, the integration of CCL leads to a substantial improvement in both the LPIPS and SNLI metrics. Visual results in Figure \ref{fig3} further demonstrate that CCL enhances CycleINR by effectively mitigating over-smoothing, as evidenced by significant reductions in LPIPS and SNLI scores.

%% file: 5_conclusion.tex
\section{Conclusion}
\label{Conclusion}
We present CycleINR, a novel 3D medical image super-resolution approach. Unlike traditional approaches, CycleINR reframes the task as function approximation, leveraging the continuity of the implicit neural representation function. This enables efficient high-resolution image generation without specific training for upsampling ratios. The introduced cycle-consistent loss mitigates over-smoothing, a local attention mechanism further enhances performance, and a novel metric, slice-wise noise level inconsistency, assesses the quality of 3D medical images. 
Overall, CycleINR shows promise for volumetric super-resolution, offering efficiency and flexibility for enhanced
visual quality, and robust downstream analysis.
For future work, extending CycleINR to diverse medical imaging modalities and evaluating its performance in larger clinical scenarios would offer insights into generalizability. Exploring applications beyond the medical field, like video interpolation, presents a compelling direction for adapting CycleINR to handle the challenges in temporal sequences.

\clearpage